# Microscopic description of Landau-Lifshitz-Gilbert type equation based on the s-d model

Akimasa SAKUMA

CREST and Department of Applied Physics, Tohoku University, Sendai, 980-8579, Japan

A Landau-Lifshitz-Gilbert type equation has been derived by using s-d model in which the s-electron system is regarded as an environment coupled weakly with the localized spins. Based on the irreducible linear response theory, we show that the relaxation function of the s-electron spin leads to the Gilbert type damping term which corresponds to the retarded resistance function in the generalized Langevin equation. The Ohmic form of the Gilbert term stems from the fact that the imaginary part of the response function (spin susceptibility) of the itinerant electron system is proportional to the frequency $\omega$ in the low $\omega$ region. It is confirmed that the Caldeira-Leggett theory based on the path-integral approach gives the same result.

§ 1 Introduction

It is of great importance, in the development of recent magnetic devices such as magnetic recording systems and the magnetic random access memory (MRAM), to acquire high speed (switching time less than 10 10 seconds) and low consumption power. In the research of MRAM, for example, in order to achieve low consumption power, exploiting current induced magnetization reversal phenomenon instead of applying external magnetic field has become a major target of study. The main subject of this item is to reduce the critical current density ($Jc$) down to order of $10^5 A/cm^2$ and much effort has been made to achieve the objective. According to Slonczewski,[1] $Jc$ is proportional to Gilbert damping coefficient $\alpha$ which appears in the Landau-Lifshitz-Gilbert (LLG) equation [2,3] widely used for studying the dynamics of magnetization. Generally, the damping term in LLG equation is believed to dominate the magnetization reversal time (usually denoted by $T_1$), and, based on the general concept of damping, it is considered to be related also to the dephasing time (usually denoted by $T_2$) of precession of spins which is currently a big concern in the development of quantum computing system. Thus there is rapidly growing interest in controlling the damping behavior to accomplish the development of magnetic devices, and accordingly the microscopic foundation of the dynamical behavior (especially of the reversal process) of magnetization is strongly desired.

The LLG equation is given by

$$\dot{M} = -\gamma M \times H_{eff} + \alpha \frac{M}{M} \times \dot{M} \quad , \tag{1}$$

where the second term in the right hand side is called Gilbert damping term, from which the reversal time ($\tau$) is expressed as $\tau \propto \alpha + 1/\alpha$ in the bulk system. Here the factor $\alpha$ is called Gilbert damping coefficient as discussed above which attracts much interest in the field of spinelectronics [4] and the recent development of magnetic devices. Besides the Gilbert damping itself, there have been many theoretical works on the microscopic origin of spin relaxation time ($\tau_s$). Elliott [5] suggested, by taking account both of the spin-orbit interaction and electron-lattice interaction, that the spin relaxation time is given by $1/\tau_s \propto (g-2)^2/\tau_R$ where $\tau_R$ denotes an electron-lattice relaxation time. The effect of spin-orbit interaction is reflected in the g-factor which deviates from 2 in the presence of orbital moment caused by the spin-orbit interaction. This form seems to be successful in explaining an experimental data[6] of temperature dependence of line-width in conduction electron resonance (CESR) spectra of several non-magnetic metals. Fulde *et al*.[7] and Singh *et al*.[8] gave a microscopic description both of spin relaxation time and spin diffusion coefficient, considering the Coulomb interaction between electrons and impurity scattering accompanying the spin flip due to spin-orbit interaction. The mechanism proposed by Kambersky[9] is a little bit different from above; he addressed the change of electronic distribution depending on the moment direction, and insisted that the phase-lag of electronic structure is reflected in the spin relaxation time. Similar mechanism was proposed by Korenman-Prange[10] with a different manner. Anyhow, all these theories rely on the spin-orbit interaction[11] in order to express the transfer of spin angular momentum to an environment system.

In the field of quantum computing system, on the other hand, intensive studies have been made on the dephasing time defined by $T_2$ in the two-state systems,[12-17] because it generally dominates the coherence time required for a computation. Since this is closely related to the time to keep the pure state (pure ensemble), the strategy is based on the evaluation of time evolution of density matrix of the relevant two-state system coupled weakly with bosonic system as an environment.[17,18] In this sense, the concept of the relaxation time of this field can be identified with what we see in the nuclear magnetic resonance (NMR). Actually, the dephasing time of spin system is usually measured by means of the NMR technique.[19,20] Theoretically, the NMR relaxation time was first studied by Korringa[21] by using the so-called s-d model which had previously been adopted by Kittel and Mitchell[22] for studying the ferromagnetic relaxation. Most recently, Sinova *et al*.[23] and Tserkovnyak *et al*.[24] also considered the s-d model to investigate the magnetization relaxation of magnetic semiconductors and conducting ferromagnets, respectively.

Though many studies have been done on the spin relaxation time as above, the microscopic understanding on the Gilbert damping especially of practical magnetic materials is still far from satisfactory level, because, in our knowledge, the direct derivation of the LLG equation from a microscopic viewpoint has not been performed. To acquire deep understanding of the various aspects of spin relaxation, we believe it is significant to inspect how the Gilbert type damping form comes about in the equation of motion. Motivated by this consideration, we aim, in the present work, mainly to derive the LLG type equation based on a microscopic model. The model used here is the s-d model in which the s-electron system is regarded as an environment coupled weakly with the localized spins. The situation is similar to that considered by Korringa [21] and Tserkovnyak *et al.* [24] who also used the s-d model. The basic stance of the present study is to build up directly the LLG equation from the s-d model and to shed light on the intrinsic effects of itinerant electron system on the spin relaxation as an open system, besides the effects of impurities and defects. The relation between the present result and that of Korringa will be discussed in section 4.

To deal with the relaxation processes in the open system, Caldeira-Leggett (CL) theory [12,25] is considered to be appropriate. The gist to be noted in the CL theory is as follows. Let the interaction between the relevant system and environment be described by linear form, and when the degree of freedom of environment system is infinite and the spectral density $J(\omega)$ is continuum, one can exclude recurrence phenomena which occur on time scales comparable to inverse level splitting of the spectrum. In this case, the long-time part of dynamics of the relevant system can appear as a relaxation process, while the on-time part can be renormalized into effective potential for relevant system as so-called counter part. In case where $J(\omega)$ is proportional to $\omega$ in the low $\omega$ region, the relaxation turns out to be Ohmic type, a type of first derivative with respect to time. We will verify that this gives the Gilbert damping term in the present case. Tatara and Fukuyama [26] have adopted the CL theory to the magnetic system, where they showed that the Stoner excitation acts as a dissipation in the depinning process of magnetic domain wall. In our case also, the damping is assigned to the Stoner excitation which is a common feature of Fermionic system.

In order to get more transparent aspect of the relaxation phenomena in the present system, we first try, before going into the CL approach, to derive the LLG equation in the language of stochastic theory based on the irreducible linear response theory. To this aim we start with the Heisenberg's equation of motion for the localized spins. It is presumed here that the dynamics of s-electron spins appearing in this equation can be expressed by the Kubo formula, by regarding the localized spins as the external field. It will be shown that the relaxation function (time-integral of the response function) of the s-electron system leads to the Gilbert damping term which corresponds to the retarded resistance function in the generalized Langevin equation. In the present case, the retarded resistance function originates from the spin susceptibility of the s-electron system. The Ohmic form of the Gilbert term stems from the fact that the imaginary part of the spin susceptibility of itinerant electron system is proportional to $\omega$ in the low $\omega$ region as is the same situation discussed above for $J(\omega)$. We will next show that the CL approach gives the same result as that obtained by the stochastic approach.

§ 2 Model

Let us consider the Hamiltonian as

$$H_M = H_S + H_e + H_{S-e} \quad , \tag{2}$$

where the first term describes localized spin system which is assumed to be in a certain ordered state. The second term represents s-electron system and the third term the interaction between the localized spins and the s-electron system, s-d interaction, which are given, respectively, by

$$H_e = -\sum_{i,j,\sigma} t_{ij} c_{i\sigma}^+ c_{j\sigma} \quad , \tag{3}$$

$$H_{S-e} = -J \sum_i \boldsymbol{S}_i \cdot \boldsymbol{\sigma}_i \quad , \tag{4}$$

$$\boldsymbol{\sigma}_i \equiv \sum_{\sigma,\sigma'} c_{i\sigma}^+ (\boldsymbol{\sigma})_{\sigma\sigma'} c_{i\sigma'} \quad . \tag{5}$$

Here $c_{k\sigma}^+ (c_{k\sigma})$ denotes creation (annihilation) operator with $\sigma \,(=\uparrow \text{or} \downarrow)$ being spin state, and $t_{ij}$ is the hopping probability amplitude between $i$-th and $j$-th sites. In eq. (4), $J$ is the exchange (Hund) coupling constant between the localized spins and s-electron spins which is assumed to be weak enough for s-electron system to be regarded as an environment. In eq. (5), $\sigma$ is Pauli matrix.

Here, for simplicity, let us suppose that the static external field $\boldsymbol{H}$ is applied to the spin at a certain site (0-th site) from time $t_0$, and the localized spin at the 0-th site is locally excited in the s-electron circumstance. This corresponds to ignoring of the interference between localized spins. Though the validity of this assumption is not guaranteed at this stage, we believe that, in an actual system, spin precession maintained both by external dc bias filed and microwave does not necessarily mean the collective mode. And then, so far as the uniform precession is concerned, the spin relaxation is assumed to take place individually. A brief discussion on the influence of interference will be given in section 4. The Hamiltonian expressing the interaction with the external field is given by $H_{ext}(t) = \gamma \boldsymbol{H} \cdot \boldsymbol{S}_0 \, \theta(t-t_0)$ ($\boldsymbol{M}_0 = -\gamma \boldsymbol{S}_0$) where $\gamma = \mu_B g / \hbar$ is the electron gyromagnetic ratio ($\mu_B$ is the Bohr magneton and $g$ is the electronic $g$ factor). Thus we have a total Hamiltonian as

$$H = H_M + H_{ext}(t) \quad , \tag{6}$$

## § 3 Derivation of a LLG type equation
### 3.1 Stochastic approach

From the Heisenberg equation of motion, dynamics of the localized spin at the 0-th site $\boldsymbol{S}_0$ is given by

$$\begin{aligned}\frac{d}{dt}\boldsymbol{S}_0(t) &= i[H, \boldsymbol{S}_0(t)] \\ &= -\gamma \boldsymbol{S}_0(t) \times \boldsymbol{H}\, \theta(t-t_0) - \gamma \boldsymbol{S}_0(t) \times \boldsymbol{H}_S + J \boldsymbol{S}_0(t) \times \boldsymbol{\sigma}_0(t)\end{aligned} \tag{7}$$

The first term in the right hand side expresses a precession of $\boldsymbol{S}_0$ around the external field $\boldsymbol{H}$. The second term represents also the precession about the effective field $\boldsymbol{H}_S$ originated from the exchange interaction between localized spins ($H_S$ in eq. (2)). Generally, this should be treated as an operator, but we assume here, for simplicity, that $\boldsymbol{H}_S$ is retained to be static field even after the external field $\boldsymbol{H}$ is applied. The third term is the precession around the s-electron spins. Provided that the localized spin is large enough to be regarded as a classical spin, one can take the expectation value of eq. (7) in terms of the electron system. Then, in this case, eq. (7) can be rewritten as

$$\frac{d}{dt}\boldsymbol{S}_0 = -\gamma \boldsymbol{S}_0 \times \{\boldsymbol{H}_S + \boldsymbol{H}\,\theta(t-t_0)\} + J \boldsymbol{S}_0 \times \langle \boldsymbol{\sigma}_0 \rangle \quad . \tag{8}$$

Worth mentioning here is that, in the absence of external field $\boldsymbol{H}$, not only the first term but also the second term of eq. (8) vanishes, since in the ground state $\langle \boldsymbol{\sigma}_0 \rangle \propto \boldsymbol{S}_0$ may be satisfied. Once $\boldsymbol{H}$ is applied to the system, $\boldsymbol{S}_0$ starts to rotate around $\boldsymbol{H}$ and then there may occur a deviation of the direction of $\boldsymbol{S}_0$ from that of $\langle \boldsymbol{\sigma}_0 \rangle$. The average of the s-electron spin $\langle \boldsymbol{\sigma}_0 \rangle$ is taken over the states of $H_e + H_{S-e}^{eq} + H_{S-e}(t)$ where we have divided eq. (4)

into static part $H_{S-e}^{eq} = -J\sum_{i}\boldsymbol{\sigma}_i \cdot \boldsymbol{S}_i^{eq}$ and the deviation from it, *i.e.* dynamical part $H_{S-e}(t) = -J\boldsymbol{\sigma}_0 \cdot \delta\boldsymbol{S}_0(t)$ which is active only at the 0-th site.

Taking advantage of linear response theory to calculate $\langle\boldsymbol{\sigma}_0(t)\rangle$, with $\delta\boldsymbol{S}_0(t)$ being regarded as an external field, we have

$$\langle\boldsymbol{\sigma}_0(t)\rangle_{H_e+H_{S-e}^{eq}+H_{S-e}(t)} = \langle\boldsymbol{\sigma}_0\rangle_{H_e+H_{S-e}^{eq}} + iJ\int_{-\infty}^{t}dt'\langle[\boldsymbol{\sigma}_0(t-t'),\boldsymbol{\sigma}_0\cdot\delta\boldsymbol{S}_0(t')]\rangle_{H_e+H_{S-e}^{eq}}$$
$$\equiv \boldsymbol{\sigma}_0^{eq} + \delta\boldsymbol{\sigma}_0(t)$$

If one adopts again the linear response theory regarding the $S_i^{eq}$ as an external field, the first term $\sigma_0^{eq}$ is shown to be expressed by the summation of RKKY interactions with surrounding spins $S_i^{eq}$. This can then be involved in the first term of eq. (8) as an effective field, and is assumed here to point to z direction. For the second term, each component can be expressed as

$$\delta\sigma_{0,x}(t) = \frac{J}{4}\{\int_{t_0}^{t}dt'\chi_{+-}(t-t')\delta S_{0,+}(t') + \int_{t_0}^{t}dt'\chi_{-+}(t-t')\delta S_{0,-}(t')\},$$

$$\delta\sigma_{0,y}(t) = -i\frac{J}{4}\{\int_{t_0}^{t}dt'\chi_{+-}(t-t')\delta S_{0,+}(t') - \int_{t_0}^{t}dt'\chi_{-+}(t-t')\delta S_{0,-}(t')\},$$

$$\delta\sigma_{0,z}(t) = J\{\int_{t_0}^{t}dt'\chi_{\uparrow\uparrow}(t-t')\delta S_{0,z}(t') + \int_{t_0}^{t}dt'\chi_{\downarrow\downarrow}(t-t')\delta S_{0,z}(t')\},$$

where we have introduced $\delta S_{\pm} = \delta S_x \pm i\delta S_y$ and $\chi_{\alpha\beta}(t) = i\langle[\sigma_\alpha(t),\sigma_\beta]\rangle\theta(t)$, $(\alpha,\beta = +,-$ or $\uparrow,\downarrow)$ with $\sigma_+ = 2c_\uparrow^+c_\downarrow$, $\sigma_- = 2c_\downarrow^+c_\uparrow$, $\sigma_\uparrow = c_\uparrow^+c_\uparrow$ and $\sigma_\downarrow = c_\downarrow^+c_\downarrow$. For $\chi_{+-}(t) = i\langle[\sigma_+(t),\sigma_-]\rangle\theta(t)$ as an example, we obtain at $T = 0$

$$\chi_{+-}(t) = -i4\theta(t)\int_{-\infty}^{\infty}d\omega\, e^{i\omega t}\int_{\varepsilon_F-\omega}^{\varepsilon_F}d\varepsilon\,\rho(\varepsilon-JS)\rho(\varepsilon+JS+\omega) \quad , \quad (9)$$

where $\rho(\varepsilon) = \sum_k \delta(\varepsilon-\varepsilon_k)$ ($\varepsilon_k = -\sum_j t_{ij}e^{-ik\cdot R_{ij}}$) is the density of states (DOS) of s-electron system and $\varepsilon_F$ the Fermi energy. The $\varepsilon$-integral in eq. (9) is just the imaginary part of spin susceptibility, $\text{Im}\chi(\omega)$ in frequency space $\omega$, corresponding to the spectral function of this system. It is known that in the Fermionic system $\text{Im}\chi(\omega)$ has a form proportional to $\omega$ in the low $\omega$ region, and goes to zero for $\omega$ larger than band width, $W$. Generally, the spectral function of environment system can be expressed with the form of $\omega^P e^{-\omega/\lambda}$ [12] where $\lambda$ is cutoff frequency, and for Fermionic system one should take $P = 1$. In the light of this theoretical picture, we approximate the $\varepsilon$-integral in eq. (9) as $\int_{\varepsilon_F-\omega}^{\varepsilon_F}d\varepsilon\,\rho(\varepsilon-JS)\rho(\varepsilon+JS+\omega) \to \rho_\uparrow\rho_\downarrow\omega e^{-|\omega|/W}$ where

$\rho_\uparrow = \rho(\varepsilon_F - JS)$ and $\rho_\downarrow = \rho(\varepsilon_F + JS)$. Thus we get

$$\chi_{+-}(t) = 8\theta(t)\rho_\uparrow \rho_\downarrow \int_0^\infty d\omega\, e^{-|\omega|/W} \omega \sin\omega t$$
$$= 8\rho_\uparrow \rho_\downarrow \phi(t) \qquad , \qquad (10)$$
$$= \chi_{-+}(t)$$

and for $\chi_{\sigma\sigma}(t) = iJ\langle[\sigma_\sigma(t), \sigma_\sigma]\rangle\theta(t)$ as well, we have

$$\chi_{\sigma\sigma}(t) = 2\rho_\sigma^2 \phi(t), \quad (\sigma = \uparrow \text{ or } \downarrow) \qquad . \qquad (11)$$

In eqs. (10) and (11), the common factor $\phi(t)$ is given by

$$\phi(t) \equiv \frac{2t/W}{[t^2 + (1/W)^2]^2}\theta(t) \qquad , \qquad (12)$$

which represents the characteristic feature of response function in Fermionic system. Eventually, the expectation value of the s-electron spin is obtained as

$$\delta\boldsymbol{\sigma}_0(t) = 4J\rho_\uparrow \rho_\downarrow \int_{t_0}^t dt'\, \phi(t-t')\delta\boldsymbol{S}_0(t') + 2J(\rho_\uparrow - \rho_\downarrow)^2 \int_{t_0}^t dt'\, \phi(t-t')\delta S_{0z}(t')\hat{z} \quad . \quad (13)$$

The occurrence of the second term in the right hand side of eq. (13) is traced back to the situation that the s-electron spins are aligned to $z$ direction in the equilibrium state in $t < t_0$ by the static part of the s-d interaction $H_{S-e}^{eq} = -J\sum_i \boldsymbol{\sigma}_i \cdot \boldsymbol{S}_i^{eq}$, and this should vanish when the s-electron system is not spin-polarized; $\rho_\uparrow = \rho_\downarrow$. As in the usual manner, the integrals in eq. (13) can be cast into the form, by integration by parts, as $\int_{t_0}^t dt'\, \phi(t-t')\boldsymbol{S}(t') =$

$\Phi(0)\boldsymbol{S}(t) - \Phi(t-t_0)\boldsymbol{S}(t_0) - \int_{t_0}^t dt'\, \Phi(t-t')\frac{d}{dt'}\boldsymbol{S}(t')$ where we have introduced the relaxation function by

$$\Phi(t) = -\int dt\, \phi(t) = \frac{1/W}{t^2 + (1/W)^2}\theta(t) \qquad . \qquad (14)$$

It should be noted that the function $\Phi(t)$ behaves as $t^{-2}$ in large $t (\gg 1/W)$ region, which is associated with the fact that $\text{Im}\chi(\omega)$ has a linear form in $\omega$ in low $\omega$ region. In the present case, the time scale of $\Phi(t)$, which is of the order of inverse of the band width, $1/W \approx 10^{-14}$ sec, is much shorter than that of the precession of localized spin, $1/\gamma H \approx 10$ sec . Thus the function $\Phi(t)$ expressed by Lorentzian type can be approximated to be delta function as a Markovian process, which gives

$$\int_{t_0}^{t} dt' \phi(t-t')\boldsymbol{S}(t') = \Phi(0)\boldsymbol{S}(t) - \Phi(t-t_0)\boldsymbol{S}(t_0) - \frac{\pi}{2}\dot{\boldsymbol{S}}(t).$$ Inserting this into eq. (13) leads

$$\delta \boldsymbol{\sigma}_0(t) = 4J \rho_\uparrow \rho_\downarrow \{\Phi(0)\delta \boldsymbol{S}_0(t) - \frac{\pi}{2}\dot{\boldsymbol{S}}_0(t)\} \\ + 2J(\rho_\uparrow - \rho_\downarrow)^2 \{\Phi(0)\delta S_{0z}(t) - \frac{\pi}{2}\dot{S}_{0z}(t')\}\hat{z} \quad , \tag{15}$$

where we exploit $\delta \boldsymbol{S}_0(t_0) = \delta S_{0z}(t_0) = 0$, and $\delta \dot{\boldsymbol{S}}_0(t) = \dot{\boldsymbol{S}}_0(t)$, $\delta \dot{S}_{0z}(t) = \dot{S}_{0z}(t)$.

Using $\delta \boldsymbol{\sigma}_0(t)$ of eq. (15), the equation of motion of the localized spin becomes

$$\frac{d}{dt}\boldsymbol{S}_0 = -\gamma \boldsymbol{S}_0 \times \boldsymbol{H}_{eff}(t) \\ - 2\pi J^2 \rho_\uparrow \rho_\downarrow \boldsymbol{S}_0(t) \times \dot{\boldsymbol{S}}_0(t) - \pi J^2(\rho_\uparrow - \rho_\downarrow)^2 \boldsymbol{S}_0(t) \times \hat{z}\dot{S}_0(t) \quad , \tag{16}$$

where the effective field $\boldsymbol{H}_{eff}(t)$ is

$$\boldsymbol{H}_{eff}(t) = \boldsymbol{H}\theta(t-t_0) + \boldsymbol{H}_S - J\boldsymbol{\sigma}_0^{eq}/\gamma \\ + 2\hat{z}J^2/\gamma\{2\rho_\uparrow \rho_\downarrow S - (\rho_\uparrow - \rho_\downarrow)^2 \delta S_{0z}(t)\}\Phi(0) \quad . \tag{17}$$

Expressing equation (16) in terms of the magnetic moment ($\boldsymbol{M} = -\gamma \boldsymbol{S}$), we reach the LLG type equation as

$$\dot{\boldsymbol{M}}_0 = -\gamma \boldsymbol{M}_0 \times \boldsymbol{H}_{eff}(t) + \alpha \frac{\boldsymbol{M}_0}{M_0} \times \dot{\boldsymbol{M}}_0 + \beta \frac{\boldsymbol{M}_0}{M_0} \times \hat{z}\dot{M}_{0z} \quad , \tag{18}$$

$$\alpha = 2\pi S J^2 \rho_\uparrow \rho_\downarrow \quad , \tag{19}$$

$$\beta = \pi S J^2 (\rho_\uparrow - \rho_\downarrow)^2 \quad . \tag{20}$$

Here we obtain the expression of the Gilbert damping coefficient $\alpha$ by eq. (19). As can be deduced from the fact that this term is traced back to the spin susceptibility of s-electron system, the occurrence of the damping, in the present case, is ascribed to the effect of retarded resistance accompanying the Stoner excitation, that is, the spin flip between ↑ spin and ↓ spin at the Fermi level. The third term in eqs. (16) and (18) does not appear in the usual LLG equation. As mentioned previously, this originates from the situation that the s-electron system stays in a spin polarized state in the equilibrium state, and is associated with the individual electron-hole pair excitations within each spin state. Since this process does not accompany the change of angular momentum, this term acts as a resistance to the precession without changing the z component of the magnetic moment.

## 3.2 Caldeira-Leggett (CL) approach

To describe relaxation processes which can be assigned to the retarded resistance due to an interaction with the environment (open) system having infinite degree of freedom, the CL approach based on the path-integral method may be adequate. The purpose of this section is to get deeper insights into the relaxation phenomenon discussed in the previous section by adopting the CL approach.

Let us consider the same situation as in the previous section. Under the Hamiltonian given by eq. (6), the action of this system is written as

$$A(S) = A_B(S) + A_e(S) + A_Z(S) + \int_0^\beta d\tau\, H_S(S) \quad , \tag{21}$$

where $A_B(S) = iS\int_0^\beta d\tau \sum_i \dot{\phi}_i(1-\cos\theta_i)$ describes the Berry phase of the localized spins, $A_Z(S) = -\int_0^\beta d\tau\, H_{ext}(\tau) = \gamma \int_0^\beta d\tau\, H \cdot S_i\, \delta_{i,0}\delta_{j,0}$ is for the Zeeman energy on the spin at the 0-th site and the last term describes the interaction between the localized spins. The second term of eq. (21) can be expressed by

$$A_e(S) = -\text{Tr}_{\omega_l,\sigma,i} \ln(-G_0^{-1}) - \text{Tr}_{\omega_l,\sigma,i} \ln(1 - G_0 M) \quad . \tag{22}$$

Here $-G_0^{-1} = -i\omega_l \hat{1} - \hat{t} - J S^{eq} \cdot \sigma$ is the Greens function of the s-electron system interacting with the localized spins in the equilibrium state, $S^{eq}$. The matrix $M$ represents the interaction of s-electrons with $\delta S_0$ which is the deviation of $S_0$ from the equilibrium state $S^{eq}$, and is written by $M_{ij} = -J\delta S_i \cdot \sigma\, \delta_{i,j}\, \delta_{i,0}$. Expanding the second term of eq. (22) up to the second order with respect to $M$, we have

$$A_e(S) = -\text{Tr}_{\omega_l,\sigma,i} \ln(-G_0^{-1}) + \text{Tr}_{\omega_l,\sigma,i}(G_0 M) + \frac{1}{2}\text{Tr}_{\omega_l,\sigma,i}(G_0 M G_0 M).$$

If we chose the z direction parallel to $S^{eq}$ ($S^{eq} = S\hat{z}$), we can write $-G_0^{-1} = -i\omega_l\hat{1} - \hat{t} - J S \sigma_z$ and obtain the first order term as

$$\text{Tr}_{\omega_l,\sigma,i}(G_0 M) = -J\int_0^\beta d\tau \delta S_{0z}(\tau) m_z \quad \text{where} \quad m_z = \frac{1}{N}\sum_k (f(\varepsilon_{k\uparrow}) - f(\varepsilon_{k\downarrow})) \quad \text{and}$$

$\varepsilon_{k\sigma} = \varepsilon_k - \sigma J S$. The second order term is thus

$$\frac{1}{2}\text{Tr}_{\omega_l,\sigma,i}(G_0 M G_0 M)$$
$$= \frac{J^2}{2N^2}\sum_l \int_0^\beta d\tau \int_0^\beta d\tau' \frac{1}{\beta} e^{i\omega_l(\tau-\tau')} \sum_{k,k'} \{(\chi_{kk'}^{\uparrow\uparrow}(i\omega_l) + \chi_{kk'}^{\downarrow\downarrow}(i\omega_l))\delta S_{0z}(\tau)\delta S_{0z}(\tau')$$
$$+ \chi_{kk'}^{\uparrow\downarrow}(i\omega_l)\delta S_{0-}(\tau)\delta S_{0+}(\tau') + \chi_{kk'}^{\downarrow\uparrow}(i\omega_l)\delta S_{0+}(\tau)\delta S_{0-}(\tau')\}$$

where $\chi_{kk'}^{\sigma\sigma'}(i\omega_l)$ is expressed by $\chi_{kk'}^{\sigma\sigma'}(i\omega_l) = \dfrac{f(\varepsilon_{k'\sigma'}) - f(\varepsilon_{k\sigma})}{i\omega_l - \varepsilon_{k\sigma} + \varepsilon_{k'\sigma'}}$. Following the manner adopted by Tatara and Fukuyama, we approximate, in the long time limit and $T \to 0$, the Fourier transformation of $\chi_{kk'}^{\sigma\sigma'}(i\omega_l)$ as

$$\sum_l e^{i\omega_l \tau} \chi_{kk'}^{\sigma\sigma'}(i\omega_l) \xrightarrow{\beta\to\infty} -\beta\delta(\varepsilon_{k\sigma} - \varepsilon_F)\delta(\varepsilon_{k'\sigma'} - \varepsilon_{k\sigma})\frac{1}{\tau^2}.$$

After some algebra, we finally obtain the second order term as

$$\frac{1}{2}\text{Tr}_{\omega_l,\sigma,i}(G_0 M G_0 M) = \frac{J^2}{2}\rho_\uparrow \rho_\downarrow \int_0^\beta d\tau \int_0^\beta d\tau' \frac{\{\mathbf{S}_0(\tau) - \mathbf{S}_0(\tau')\}^2}{(\tau-\tau')^2}$$
$$+ \frac{J^2}{4}(\rho_\uparrow - \rho_\downarrow)^2 \int_0^\beta d\tau \int_0^\beta d\tau' \frac{\{S_{0Z}(\tau) - S_{0Z}(\tau')\}^2}{(\tau-\tau')^2} \quad . \quad (23)$$
$$- J^2 \int_0^\beta d\tau[\rho_\uparrow\rho_\downarrow \delta \mathbf{S}_0^2(\tau) + \frac{1}{2}(\rho_\uparrow - \rho_\downarrow)^2 \delta S_{0Z}^2(\tau)]\int_0^\beta d\tau' \frac{1}{(\tau-\tau')^2}$$

In the derivation of eq. (23), we take advantage of the CL technique given by $2\delta S(\tau)\delta S(\tau') = -\{S(\tau) - S(\tau')\}^2 + \delta S^2(\tau) + \delta S^2(\tau')$ where the subscripts are dropped for brevity. This decomposition enables us to extract the non-local part in terms of the time, which brings about the first two terms in eq. (23). As shown below, the non-local terms express dissipation as a long time behavior of the system, while the local terms can be included in the effective potential as an effective field.

The equation of motion for $\mathbf{S}_0$ can be obtained by a variation of action $A(S)$ with respect to $\mathbf{S}_0$, $\delta A/\delta \mathbf{S}_0 = 0$. Then we have

$$\gamma H_{eff}(\tau) + i \dot{S}_0(\tau) \times S_0(\tau) + 2J^2 \rho_\uparrow \rho_\downarrow \int_0^\beta d\tau' \frac{(S_0(\tau) - S_0(\tau'))}{(\tau - \tau')^2}$$
$$+ J^2 (\rho_\uparrow - \rho_\downarrow)^2 \int_0^\beta d\tau' \frac{(S_{0z}(\tau) - S_{0z}(\tau'))z}{(\tau - \tau')^2} = 0 \quad , \quad (24)$$

where the effective field is given by

$$H_{eff}(\tau) = H \delta_{i,0} + H_S - J m_z \hat{z}/\gamma$$
$$+ \hat{z} J^2/\gamma \{2\rho_\uparrow \rho_\uparrow S - (\rho_\uparrow - \rho_\downarrow)^2 \delta S_{0z}(\tau)\} \int_0^\beta d\tau \frac{1}{(\tau - \tau')^2} \quad . \quad (25)$$

One can notice that $H_{eff}(\tau)$ of eq. (25) exhibits a good correspondence with that given by eq. (17) where $\int_0^\beta d\tau (\tau - \tau')^{-2}$ in eq. (25) is replace by $2\Phi(0)$.

The factor $\int_0^\beta d\tau (\tau - \tau')^{-2}$ which goes infinite for $|\tau - \tau'| \to 0$ is ascribed to the long time approximation, so it should remain finite in principle. As seen in eq. (14), $\Phi(0)$ has the order of band width $W$ of the s-electron system, which can be approximated to be inverse of the DOS, $\rho^{-1}$. Thus the last term in eq. (25) can be inferred to give the contribution of around $J^2 \rho/\gamma$, the same or smaller order as the third term in eq. (25).

In the scenario of the CL theory, the form like the third and forth terms of eq. (24) has the contribution of the first derivative with respect to time, that is, Ohmic type dissipation. Mathematically, however, these two terms are a little bit different from the typical form of the CL theory. Therefore some approximations have been made in this work. The details are given in Appendix. Eventually, by multiplying $S_0$ to eq. (24) from the left and letting $\tau \to it$, the equation of motion is recast into the form of

$$\dot{S}_0 = -\gamma S_0 \times H_{eff} - \pi J^2 (2\rho_\uparrow \rho_\downarrow (S_0 \times \dot{S}_0) + (\rho_\uparrow - \rho_\downarrow)^2 \dot{S}_{0z} S_0 \times z) \quad . \quad (26)$$

Thus we can reach the same result as eq. (16) obtained by the stochastic approach. In both approaches, the Gilbert damping term stems from the imaginary part of spin susceptibility of Fermionic system which behaves as $\omega$ linear in low $\omega$ region. Worth mentioning here is that the stochastic theory based on the dissipation-fluctuation theory supplies substantially the same result with the CL approach based on the path-integral theory. The comparison of these two approaches tells us that the decomposition of susceptibility into the long time part and the counter part in the CL theory corresponds to, in the stochastic approach, doing the integration by parts for the response function term to obtain the relaxation function. And the integrated value of the relaxation function turns out to correspond to the Gilbert damping coefficient $\alpha$, which is just the long time part of the spin susceptibility in the CL approach. It should be noted that, as can be understood from eq. (14), the relaxation function corresponding to the Gilbert damping coefficient is time-dependent, in principle, which should appear with a form of $\int_{t_0}^{t} dt' \Phi(t-t') \dot{S}(t')$ in the LLG equation. The extremely slow motion of spin precession compared with the relaxation of the s-electron system enable us to treat this term as $\dot{S}(t) \int_{t_0}^{t} dt' \Phi(t-t') \propto \alpha \dot{S}(t)$, as a Markovian process. This feature is manifested as a long time part in the CL approach.

4. Discussion

We have derived the LLG type equation directly from the s-d model and showed that the Gilbert damping $\alpha$ is proportional to the DOS of each spin state. Here, based on the present result, let us make briefly a quantitative analysis of the Gilbert damping coefficient of permalloy as a typical transition metal system. The s-electron system in the present model can be regarded as sp bands in the transition metal system and the localized spin system as the d-electron system. From the first principles band calculation combined with the coherent potential approximation (CPA), the DOS of sp bands at the Fermi level in permalloy are found to be around 0.04 eV$^{-1}$ per atom and spin for both spin states and one can put $2S=1$. If we take $J=1$eV, we obtain $\alpha$ of about 0.003 which is of the same order as the measured value, $\alpha \approx 0.008$.[27] From the present result we expect that in the half-metallic system where the DOS of either spin state vanishes the Gilbert damping is expected to be zero. And, in fact, in YIG which has no Fermi surface in both spin states the damping is known to be quite small.

The similar model and result as ours were previously presented by Korringa who discussed the nuclear spin relaxation time $T_1$ by taking into account the interaction with the conduction electrons. He demonstrated that the relaxation rate $1/T_1$ is proportional to both the DOS and temperature $T$ through

$$\sum_{k,k'} f(\varepsilon_k)(1-f(\varepsilon_{k'}))\delta(\varepsilon_k - \varepsilon_{k'}) = -k_B T \sum_{k,k'} \frac{\partial f(\varepsilon_k)}{\partial \varepsilon_k}\delta(\varepsilon_k - \varepsilon_{k'}) \cong k_B T \rho^2.$$

This is similar to our result except for the proportionality to $T$. Mathematically, this discrepancy is naturally understood by looking that the Korringa relaxation term given above can be expressed by $\frac{k_B T}{\omega}\operatorname{Im}\chi(\omega)$, while the response function in our case is proportional to $\operatorname{Im}\chi(\omega)$ itself. Physically, the Korringa relaxation provides the transition probability of localized spin through the interaction with the conduction electrons, which remains finite even for $\omega=0$, with a help of thermal excitation of the conduction electrons. On the other hand, $\operatorname{Im}\chi(\omega)$ represents the spin fluctuation spectrum of itinerant electron system in the equilibrium state and also describes the response to the external field having frequency of $\omega$. In the Fermionic system this behaves as linear in $\omega$. Thus it must vanish for $\omega \to 0$ even in finite temperatures. These differences are reflected in the difference of damping form in the equation of motion, that is, $(M-M_z)/T_1$ for Korring case and $M \times \dot{M}$ for ours. As a physical property, the difference would appear clearly in the intrinsic temperature dependence of the damping constant. However, looking at the experimental data of transition metal systems obtained by Bhagat *et al.* [28], the temperature dependence of damping constant seems not so simple nor systematic, which may imply that the mechanism of spin relaxation in transition metal systems has various aspects depending on the temperature, as suggested by Kambersky. [29]

Finally, let us make a brief consideration on the interference effects of localized spins. If we extend our stochastic approach to treat the collective mode of spin precession, the equation of motion can be rewritten in the form of

$$\dot{\boldsymbol{S}}_i(t) = -\gamma\,\boldsymbol{S}_i(t) \times \boldsymbol{H}_{eff} + J\,\boldsymbol{S}_i(t) \times \sum_j \int_{t_0}^{\infty} dt'\,\chi_{ij}(t-t')\delta\boldsymbol{S}_j(t').$$

Here $\chi_{ij}(t)$ corresponds to the non-local spin susceptibility whose Fourier transformation is given by $\chi(\boldsymbol{q},\omega)$. The imaginary part of $\chi(\boldsymbol{q},\omega)$ generally has the form of $\omega/q$ which also behaves as $\omega$ linear at low $\omega$ region. This should vanish in the limit $\boldsymbol{q} \to 0$, which implies that the uniform precession, so called the Kittel mode, does not have finite Gilbert damping. In fact, according to the theoretical result presented by Singh *et al.*,[8] the damping of spin wave due to the spin flip scattering is proportional to $q^2$ which goes to zero for $q$=0. However, almost experimental data show that in a uniform precession the Gilbert damping remains finite. It should be noted here that, in actual experimental situations, the uniform precession is retained compulsorily with a help of external microwave. This situation leads us to expect that the uniform precession in the actual system does not necessarily reflect the collective mode of $q$=0 but can be considered as an individual precession with no interference. For NMR relaxation, at least, the interpretation of spin relaxation in terms of local precession seems to work well in most experimental data. A detailed analysis will be given in the next paper.

In summary, we have derived directly the LLG type equation using the s-d model in which the s-electron system is regarded as an environment coupled weakly to the localized spins. Based on the irreducible linear response theory, we show that the relaxation function (time-integral of the response function) of the s-electron system leads to the Gilbert damping term which corresponds to the retarded resistance function in the generalized Langevin equation. The Ohmic form of the Gilbert term stems from the fact that the imaginary part of the response function corresponding to the spin susceptibility of itinerant electron system is proportional to $\omega$ in the low $\omega$ region, being ascribed to the Stoner excitation. The new finding is the additional term which behaves as $\dot{M}_{0z} M_0 \times \hat{z}$, originating from the individual electron-hole pair excitations in each spin sate. Further we confirm that this approach supplies substantially the same result as that by Caldeira-Leggett theory based on the path-integral approach.

The author thanks to Profs. S. Maekawa, H. Imamura and H. Tsuchiura for helpful discussion.

**Appendix - derivation of damping term in eq. (26) -**

Let us consider a function $F(\tau)$ defined below.

$$F(\tau) \equiv \int_0^\beta d\tau' \frac{(M(\tau) - M(\tau'))}{(\tau - \tau')^2}$$

$$= \int_0^\beta d\tau' \int_0^\infty d\omega\, e^{-\omega|\tau-\tau'|} \omega\, (M(\tau) - M(\tau'))$$

(A-1)

**Making use of the approximation**

$$\frac{2}{\beta} \sum_m e^{i\omega_m \tau} \frac{\omega^2}{\omega_m^2 + \omega^2} = \frac{e^{\omega|\tau|}}{e^{\beta\omega} - 1} + \frac{-e^{-\omega|\tau|}}{e^{-\beta\omega} - 1} \xrightarrow[\tau<\beta\to\infty]{} e^{-\omega|\tau|},$$

(A-2)

$F(\tau)$ **can be put into a form which is valid for** $T = 0$

$$F(\tau) \to \frac{1}{\beta} \int_0^\beta d\tau' \int_0^\infty d\omega \sum_m e^{i\omega_m(\tau-\tau')} \frac{2\omega^2}{\omega_m^2 + \omega^2} (M(\tau) - M(\tau'))$$

$$= \frac{1}{\beta} \sum_l M(i\omega_l) e^{-i\omega_l \tau} \int_0^\infty d\omega \frac{2\omega_l^2}{\omega_l^2 + \omega^2}$$

(A-3)

**If we put** $i\omega_l \to \varepsilon + i\delta$, **then**

$$\int_0^\infty d\omega \frac{2\omega_l^2}{\omega_l^2 + \omega^2} \to \varepsilon \int_0^\infty d\omega \left(\frac{1}{\omega + \varepsilon + i\delta} - \frac{1}{\omega - \varepsilon - i\delta}\right) = -i\pi\varepsilon$$

(A-4)

**Replacing the frequency summation in eq. (A-3) by integral in terms of** $\varepsilon$ **and putting** $\tau \to it$, **one finally obtains**

$$F(\tau) \to \frac{1}{\beta} \sum_l M(i\omega_l) e^{-i\omega_l \tau} \int_0^\infty d\omega \frac{2\omega_l^2}{\omega_l^2 + \omega^2}$$

$$\to \pi \int_{-\infty}^\infty d\varepsilon\, M(\varepsilon) e^{-i\varepsilon t}(-i\varepsilon)$$

$$= \pi \dot{M}(t)$$

(A-5)